\documentclass[sigconf]{acmart}
\usepackage{booktabs}
\usepackage{multirow}
\usepackage{graphicx}
\usepackage{xcolor}
\usepackage{tikz}
\usetikzlibrary{shapes.geometric, arrows.meta, positioning, calc}

\copyrightyear{2026}
\acmYear{2026}
\setcopyright{cc}
\setcctype{by}
\acmConference[VRST '26]{32nd ACM Symposium on Virtual Reality Software and Technology}{November 16--18, 2026}{Sendai, Japan}
\acmBooktitle{32nd ACM Symposium on Virtual Reality Software and Technology (VRST '26), November 16--18, 2026, Sendai, Japan}
\acmDOI{10.1145/3822517.3848720}
\acmISBN{979-8-4007-2811-2/2026/11}

\tikzset{
  pbox/.style   ={rectangle, draw, rounded corners=2pt, align=center,
                  text width=42mm, inner sep=4pt, font=\footnotesize, minimum height=10mm},
  pexcl/.style  ={rectangle, draw, rounded corners=2pt, align=left,
                  text width=40mm, inner sep=4pt, font=\scriptsize, minimum height=10mm},
  parrow/.style ={-{Latex[length=2mm]}, thick}
}

\begin{document}

\title[Game-Based versus Training-Based VR for Cognitive Rehabilitation]
{Game-Based versus Training-Based Virtual Reality for Cognitive Rehabilitation in Mild Cognitive Impairment and Dementia: A Systematization of Design Paradigms, Outcomes, and Evaluation Rigor}

\author{Anowarul Faruk Shishir}
\affiliation{%
  \department{Dept. of Computer Science}
  \institution{Kennesaw State University}
  \city{Marietta}
  \state{Georgia}
  \country{USA}
}
\email{mshishir@students.kennesaw.edu}

\author{M. Rasel Mahmud}
\affiliation{%
  \position{Dept. of Computer Science}
  \institution{Kennesaw State University}
  \city{Marietta}
  \state{Georgia}
  \country{USA}
}
\email{mmahmud2@kennesaw.edu}

\begin{abstract}
Therapeutic exercises play a crucial role in recovering from or slowing the cognitive decline in dementia. However, these processes are tedious and many patients give up before achieving desired benefits. Virtual Reality has evolved as an alternative to address this issue and it has two different paradigms in the rehabilitation process. The first one is a game-based approach that is designed to keep the participants engaged through enjoyable interactions. The second one provides a structured environment intended to facilitate functional recovery. A few prior studies have compared these paradigms in terms of clinical outcomes. The findings in the previous reviews are inconsistent. In addition, whether these labels identify systems that actually differ in design or delivery remains unexplored. Furthermore, whether these technological differences contribute to the inconsistencies found in the literature leaves an open question which should be investigated. This study addresses this gap in the literature. Following PRISMA 2020 guidelines, we systematize 34 studies published between 2015 and 2026, including 16 controlled efficacy trials and 18 design and feasibility studies focused on mild cognitive impairment, dementia and Alzheimer's disease. We initially built a taxonomy to classify each system according to design paradigm, immersion level, feedback modality, difficulty adaptation mechanisms and hardware. Our findings indicate that both game-based and training-based approaches exhibit some common characteristics, which made it challenging to clearly differentiate these paradigms on technological grounds. We further develop an evaluation framework, VR-RES, and apply it across existing studies. This framework highlights the limitations of previous work, particularly the tendency to neglect demographic factors such as age, sex, and education when analyzing the effectiveness of these paradigms. Finally, we outline a set of recommendations to guide future study design, evaluation and reporting practice.

\end{abstract}

\keywords{Virtual Reality, Cognitive Rehabilitation, Mild Cognitive Impairment,
Dementia, Alzheimer's Disease, Serious Games, Systematization of Knowledge, PRISMA}

\ccsdesc[500]{Human-centered computing~Virtual reality}
\ccsdesc[300]{Applied computing~Health informatics}

\maketitle
\section{Introduction}
\label{sec:intro}
Age-related neurodegenerative conditions are one of the major evolving issues in public health nowadays. The number of people living with dementia is rising enormously and projected to rise from 57 million in 2019 to 153 million by 2050~\cite{gbd2022dementia}. In addition to that, many older people live with mild cognitive impairment (MCI), an early stage of dementia, which is also characterized by a gradual decline of cognitive abilities. Since this is the early stage, interventions have the potential to slow the decline at this stage~\cite{petersen2004}. So far, pharmacological treatments have shown very little improvement. As a result, non-pharmacological interventions can play a vital role in patient care. Recent studies suggest that cognitive training helps older adults slow down the cognitive decline~\cite{livingston2020}. However, people often do not stick with these programs. Participants lose interest over time as traditional paper-based or computer programs are unengaging. 

Virtual reality (VR) has become an alternative tool for cognitive rehabilitation~\cite{hossain2025xrreview} because it can make training immersive and realistic. As a result, users can feel a sense of presence and do not get bored. Currently, two types of VR systems are available for cognitive rehabilitation. One is a game-based system that uses rewards, stories and play to keep the user motivated and engaged~\cite{slater2009}. Another is a training-based system that is focused on structured and clinically grounded exercises. The division between motivation and protocol affects design and how trials compare them.

There are some inconsistencies in the current literature about which paradigm works better. Some previous literature reported the game-based format as the better option~\cite{yuan2025review, moulaei2024review}. In contrast, other studies did not identify either training-based or game-based systems as the better option~\cite{yang2025jmirreview, santos2025review}. However, these reviews treat the two labels as different without checking whether the systems differ in any codeable feature. Before asking which approach is more effective, we should establish whether the labels identify distinguishable systems at all.

Here, we address this question directly. We systematized 34 VR cognitive rehabilitation papers across 5 design features. We examined whether these two paradigms are differentiable in design or delivery. Our findings suggest that the split did not hold. There was a substantial amount of overlap in all features, including immersion, feedback, difficulty adaptation and hardware between these two paradigms. This result challenges the fundamental assumption of considering both interventions as different. Since they are not separable, the inconsistencies found in earlier studies cannot be explained by their paradigm label alone. The differences in findings might come from their design features and study methodologies. To investigate this further, we also review the cognitive outcomes reported in controlled studies and evaluate the quality of the available evidence.
This paper makes the following contributions:
\begin{itemize}
\item  We build a multi-axis taxonomy to classify VR cognitive-rehabilitation systems by paradigm, immersion, feedback, difficulty adaptation, and hardware. We show that the paradigm label does not predict a system's technological configuration.
\item We then summarize the cognitive outcomes from 16 controlled studies and explore whether they differ across intervention paradigms.
\item Finally, we introduce VR-RES, which is a seven-dimensional framework for assessing study quality. It reveals the common weaknesses of prior studies and recommends the factors to be considered in future studies.
\end{itemize}

\section{Background and Related Work}
\label{sec:related}
\subsection{VR for cognitive rehabilitation:}
The implementation of VR in cognitive rehabilitation comes with a lot of advantages. Our everyday activities such as shopping, cooking and navigation can be better simulated by VR technology. This makes VR more realistic and suitable compared to traditional paper-based tasks. Therefore, training and assessment of different cognitive conditions can be done more effectively through VR. In addition, immersive environments help users feel present, which can improve motivation and engagement in training~\cite{slater2009}. The performance can be monitored in detail through VR technology~\cite{pavel2025patchfusionvr, pavel2025vrfallnet, pavel2026digitaltwin}. It also comes with the advantages of adjusting task difficulty in real time and delivering feedback through multiple sensory channels~\cite{mahmud2023feedbackmodalities, mahmud2025vibrotactilewalking, mahmud2026multimodalwalking}. Together, these features enhance the engagement and make the training more realistic. Therefore, VR is considered a more effective tool compared to conventional cognitive training~\cite{yuan2025review, gao2025review}. Although a few prior works have investigated VR-based interventions for balance and gait \cite {9995441,9756779,mahmud2022vibrotactile,mahmud2023multimodal,mahmud2022standing,mahmud2023eyes,mahmud2023visual,mahmud2023auditory,mahmud2024multimodal,cordova2023real, hossain2026cogniaudit,hossain2025too}, comparatively fewer studies have focused specifically on VR-based cognitive rehabilitation.

\subsection{Game-based and Training-based Paradigms:}
According to prior studies, systems are usually classified into two groups. The first approach is a game-based approach that includes serious games and gamified exercises. This intervention uses points, levels, stories and rewards to keep the users motivated. The enjoyment users receive keeps them engaged with the tasks required for cognitive gains. The second approach is a training-based approach that focuses on clinical accuracy. Here tasks are more connected to specific skills such as memory, attention and executive function. The differences arise from the intent. One prioritizes enjoyment. The other strictly follows the protocol. Researchers use this dissimilarity to organize design and structure comparisons~\cite{moulaei2024review, yang2025jmirreview}. 
\subsection{Category assessment in previous studies:}
A good number of works have worked with this evidence, though there is no agreement between them. Yuan et al.~\cite{yuan2025review} and Moulaei et al.~\cite{moulaei2024review} mentioned that game-based intervention provides higher cognitive benefits compared to the other. On the contrary, other reviews found no clear advantage for any approach. Instead, they indicated different factors such as dose, population and comparator that contribute to the improvements~\cite{yang2025jmirreview,
santos2025review, gao2025review}. Reviews that are focused on motor and balance outcomes have different findings, where results depend on the scope and method of the review~\cite{rodriguezmansilla2025review}.
Many previous works compare the outcomes between game-based and training-based interventions without checking whether they are actually different systems. It is not clear if game-based and training-based labels point to design differences that can explain the outcomes. Before comparing the effectiveness, it is very important to evaluate the validity of the categories. If the labels do not refer to real design differences, then the comparison does not yield meaningful results. Therefore, the advantage could be related to specific design features or study methods, not to the intervention itself. Consequently, we first validate the categories using a taxonomy and complete a statistical analysis of paradigm separation (\S~\ref{sec:taxonomy}). Following that, we review the cognitive outcomes (\S~\ref{sec:outcomes}) and assess how rigorously the studies were conducted (\S~\ref{sec:vrres-results}). We test category validity first to determine whether the game-based vs training-based division is a useful way to organize the field. 

\section{Research Questions and Methodology}
\label{sec:method}

\subsection{Research Questions}
\label{sec:rqs}
This systematization is organized around three research questions:
\begin{description}
\item[RQ1 (Design).] How can VR Cognitive-Rehabilitation Systems be described in terms of paradigm, immersion, feedback, difficulty adaptation and hardware for MCI, dementia and AD? Are game-based and training-based systems different in terms of their technological characteristics?
\item[RQ2 (Outcomes).] What cognitive outcomes are reported in controlled VR interventions? Are certain choices associated with greater effectiveness?
\item[RQ3 (Rigor and reporting).] How strong is the current evidence base? What reporting gaps make results hard to interpret?
\end{description}

\subsection{Methods for Systematization}
\label{sec:methods}
\subsubsection{Protocol: }
\begin{figure}[t]
\centering
\tikzset{pbox/.append style={text width=36mm}, pexcl/.append style={text width=34mm}}
\begin{tikzpicture}[node distance=8mm and 7mm]

\node[pbox] (ident)
  {Records identified from databases ($n=3{,}166$):\\
   PubMed $1{,}073$; IEEE~Xplore $356$; ACM~DL $37$;\\ Scopus/WoS $1{,}700$};
\node[pexcl, right=of ident] (dup)
  {Duplicate records removed $=935$\\
   {\tiny(1{,}819 flagged; 884 retained after review)}};

\node[pbox, below=of ident] (screen)
  {Records screened (title/abstract)\\ $n=2{,}231$};
\node[pexcl, right=of screen] (screxcl)
  {Records excluded $=2{,}132$};

\node[pbox, below=of screen] (assessed)
  {Reports assessed for eligibility\\ $n=99$};
\node[pexcl, right=of assessed] (eligexcl)
  {\quad Excluded with reasons $=62$:\\ 
   \quad Wrong population $\approx21$\\
   \quad Motor/physical-only $\approx10$\\
   \quad Conceptual/protocol/review $\approx9$\\
   \quad Single-arm, no comparator $\approx8$\\
   \quad Not retrievable $=9$\\
   \quad Non-English / not VR / other $\approx5$\\
   \quad Companion reports linked $=3$};

\node[pbox, below=of assessed, fill=black!4] (incl)
  {Studies included in synthesis\\ $n=34$\\
   Tier~1 (controlled) $=16$;\ Tier~2 (design) $=18$};

\draw[parrow] (ident) -- (dup);
\draw[parrow] (ident) -- (screen);
\draw[parrow] (screen) -- (screxcl);
\draw[parrow] (screen) -- (assessed);
\draw[parrow] (assessed) -- (eligexcl);
\draw[parrow] (assessed) -- (incl);

\end{tikzpicture}
\caption{PRISMA 2020 flow of study selection.}
\label{fig:prisma}
\end{figure}
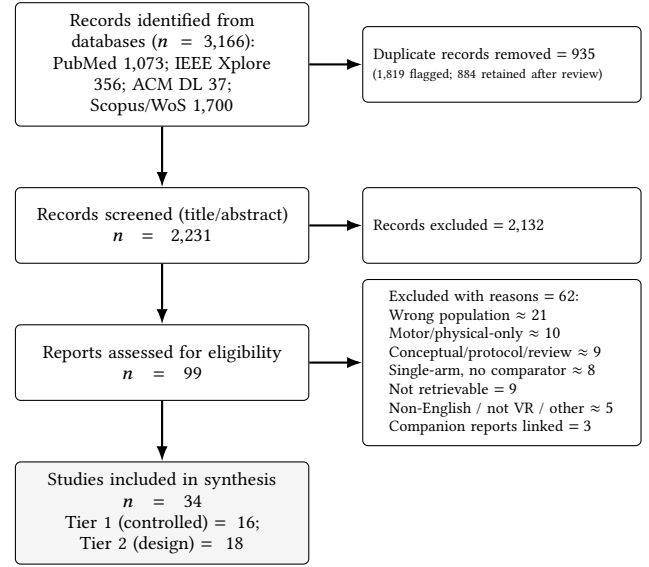

This review was completed following the PRISMA 2020 guidelines~\cite{page2021prisma}. The initial search scope was intentionally large. It included VR rehabilitation studies of both motor and cognitive functions. Hence, the search terms also included topics like stroke, Parkinson's disease, balance and gait (Table~\ref{tab:search}). During the title and abstract screening, we narrowed our scope to cognitive rehabilitation for age-related conditions, mild cognitive impairment (MCI), dementia and Alzheimer's disease (AD). Studies that focused solely on motor or physical rehabilitation were excluded. However, studies that included both motor and cognitive rehabilitation were kept in our study for further analysis. This change in focus is reported transparently in our PRISMA flow diagram (Figure~\ref{fig:prisma}).

\subsubsection{Search Strategy and Information Extraction: }We searched 4 databases. They are PubMed, IEEE Xplore, ACM Digital Library and Scopus/Web of Science. We targeted the studies that were published between 2015 and 2026 in the English language. The search keywords include three groups of keywords: (1) Virtual Reality, (2) both interventions (Game-based and Training-based) and (3) target populations. A total of 3,166 records were retrieved from the four databases. The number of records identified from each database is presented in Table~\ref{tab:search}.

\begin{table*}[t]
\caption{Search strategy and per-database results.}
\label{tab:search}
\small
\begin{tabular}{@{}p{0.16\textwidth}p{0.66\textwidth}r@{}}
\toprule
\textbf{Database} & \textbf{Exact search string used} & \textbf{Results} \\
\midrule
PubMed / MEDLINE &
(``virtual reality'' OR VR OR ``head-mounted display'' OR immersive) AND
(``serious game'' OR exergame OR ``game-based'' OR gamified OR ``cognitive
training'' OR ``task-oriented'' OR ``motor training'') AND (rehabilitation OR
stroke OR Parkinson OR ``multiple sclerosis'' OR ``cognitive impairment'' OR
dementia OR Alzheimer OR balance OR gait) & 1{,}073 \\
\addlinespace
IEEE Xplore &
Same three concept blocks combined with AND, entered as three All-Metadata rows:
(VR terms) AND (game/training terms) AND (rehabilitation/condition terms). & 356 \\
\addlinespace
ACM Digital Library & Identical Boolean string (syntax-adapted to the platform). & 37 \\
\addlinespace
Scopus / Web of Science & Identical Boolean string (syntax-adapted to the platform). & 1{,}700 \\
\midrule
\textbf{Total identified} & & \textbf{3{,}166} \\
\bottomrule
\end{tabular}
\par\smallskip
{\footnotesize\raggedright The same Boolean string was used across PubMed, the ACM
Digital Library, and Scopus/Web of Science, with minor syntactic adaptation per
platform; IEEE~Xplore used the same three concept blocks entered as separate
All-Metadata rows.\par}
\end{table*}

\subsubsection{Eligibility Criteria:} We used a two-tier inclusion framework (Table~\ref{tab:criteria}). Tier-1 studies were used to evaluate the effectiveness of the interventions. In contrast, Tier-2 studies provide information about the system design, but they do not cover information to evaluate the effectiveness of the systems.

\begin{table}[t]
\caption{Two-tier eligibility criteria and exclusion grounds.}
\label{tab:criteria}
\small
\begin{tabular}{@{}p{0.18\columnwidth}p{0.74\columnwidth}@{}}
\toprule
\textbf{Tier} & \textbf{Definition} \\
\midrule
Tier~1 \newline (Synthesis) &
RCT or controlled non-randomized study; VR game-based, training-based, or hybrid
cognitive programs; MCI/dementia/AD population; at least 1 measurable cognitive
outcome. Full taxonomy encodable, outcome extractable, and VR-RES scoring supported;
feeds RQ2 and RQ3. \\[2pt]
Tier~2 \newline (Design) &
Feasibility, usability, or acceptability study of a VR cognitive system
that can be coded, in the MCI/dementia/AD population. No controlled efficacy outcome. Receives
taxonomy coding only; supports RQ1 and system origin. \\
\midrule
\multicolumn{2}{@{}p{0.95\columnwidth}@{}}{\textbf{Excluded:} single-arm
pre--post efficacy pilots with no comparator; motor/physical-only exergames with
no cognitive outcome; conceptual, perspective, or protocol papers
reporting no data; wrong population (e.g.\ stroke, traumatic brain injury,
psychiatric, healthy young adults); augmented or mixed-reality systems; records outside 2015--2026 or not in English.}\\
\bottomrule
\end{tabular}
\end{table}

\subsubsection{Screening and Scope Evaluation: }After importing 3166 records, 1819 studies were identified as duplicates. We then went for manual verification and removed 935 duplicate records. We retrieved 884 studies from the duplicates and therefore found a total of 2231 unique records for title and abstract screening. At this stage, we eliminated 2132 records since they were either outside the scope of MCI, dementia and AD or they were solely focused on motor rehabilitation. Hence, 99 studies went through the full text screening stage. We excluded 62 studies here with reasons and 3 other studies were reported as companion publications, which we already included in our study. Only 34 studies satisfied all our criteria, and they were divided into 2 Tiers (16 Tier 1 and 18 Tier 2). The complete screening process is illustrated in Figure~\ref{fig:prisma}.

\subsubsection{Data Extraction and Classification: }Each included study was described using 5 design features that include intervention type, level of immersion, feedback modality, adaptation and hardware. We also considered population characteristics and targeted cognitive domains to describe these studies. Additional information such as study design, sample size, comparator and cognitive outcomes was also considered for Tier-1 studies to address RQ2.

\subsubsection{Quality Assessment: }To assess the quality of the selected studies, we introduced a framework, VR-RES (VR Rehabilitation Evidence Score). This is divided into 7 dimensions, where each dimension is scored from 0 to 2. Hence, the maximum score of evaluation is 14. Only the Tier-1 studies were evaluated using this framework. The framework is explained in detail in \S\ref{sec:vrres-results} along with the evaluation results.

\subsubsection{Synthesis Method: }Since there was variety in the outcome measurement of the studies, a meta-analysis was not feasible. We, therefore, went for the structured analysis. For RQ1, we looked at the distribution of taxonomy dimensions. We also evaluated whether there is any difference in the design features of game-based, training-based and hybrid systems. The direction and magnitude of cognitive effects were evaluated in Tier-1 studies for RQ2. To address RQ3, we analyzed the VR-RES scores and checked how demographic moderators were reported. 

\section{Taxonomy of VR Cognitive-Rehabilitation Systems (RQ1)}
\label{sec:taxonomy}
We coded all 34 systems using five design and technology dimensions which are introduced in \S~\ref{sec:methods}. The results are summarized in Table~\ref{tab:taxonomy}. From the analysis, two observations become evident. First of all, there are no dominant features such as immersion, hardware type, feedback strategy across the literature. Secondly, there are similarities in the design and implementation of game-based and training-based systems. Therefore, they should not be considered as technologically distinct system classes.

\begin{table*}[t]
\caption{Taxonomy of the 34 included VR cognitive-rehabilitation systems.}
\label{tab:taxonomy}
\footnotesize
\begin{tabular}{@{}lccccccl@{}}
\toprule
\textbf{Study} & \textbf{T} & \textbf{Par} & \textbf{Imm} & \textbf{FB} & \textbf{Adj} & \textbf{HW} & \textbf{Pop} \\
\midrule
Kwan et al. 2024~\cite{kwan2024} & 1 & H & F & MM & M & HMD & MCI/Dem/Frailty \\
Liao et al. 2020~\cite{liao2020} & 1 & T & F & MM & M & HMD & MCI \\
Buele et al. 2024~\cite{buele2024} & 1 & T & F & MM & A & HMD & MCI \\
Baldimtsi et al. 2023~\cite{baldimtsi2023} & 1 & H & F & MM & M & HMD & MCI \\
Hassandra et al. 2021~\cite{hassandra2021} & 2 & H & F & MM & M & HMD & MCI \\
Manera et al. 2016~\cite{manera2016} & 2 & T & S & V & M & Proj & MCI/Dem \\
Yun et al. 2020~\cite{yun2020} & 2 & H & F & MM & M & HMD & MCI/Dem \\
Maeng et al. 2021~\cite{maeng2021} & 2 & T & F & MM & M & HMD & MCI \\
Na et al. 2025~\cite{na2025} & 2 & G & F & MM & A & HMD & MCI \\
Mondellini et al. 2022~\cite{mondellini2022} & 2 & T & F & MM & Fx & HMD & MCI \\
Ip et al. 2025~\cite{ip2025} & 1 & H & S & MM & M & CAVE & MCI \\
Mrakic-Sposta et al. 2018~\cite{mrakicsposta2018} & 1 & H & N & MM & A & Proj & MCI/Dem \\
Hsieh et al. 2018~\cite{hsieh2018} & 1 & H & N & MM & A & Kinect & MCI/Dem \\
Liao et al. 2019~\cite{liao2019} & 1 & H & N & MM & A & HMD & MCI \\
Zhu et al. 2022~\cite{zhu2022} & 2 & T & F & MM & A & HMD & MCI/Dem \\
Thapa et al. 2020~\cite{thapa2020} & 1 & G & F & MM & Fx & HMD & MCI \\
Latella et al. 2024~\cite{latella2024} & 2 & T & N & MM & Fx & Tablet & MCI \\
Kua et al. 2025~\cite{kua2025} & 1 & T & F & MM & M & HMD & MCI \\
Park 2022~\cite{park2022} & 1 & T & N & V & Fx & Desktop & MCI \\
Yang et al. 2022~\cite{yang2022} & 1 & H & F & MM & A & HMD & MCI \\
Zuo et al. 2024~\cite{zuo2024} & 2 & G & F & MM & A & HMD & AD \\
Rings et al. 2020~\cite{rings2020} & 2 & H & F & MM & M & HMD & Dem \\
Tuena et al. 2022~\cite{tuena2022} & 2 & T & S & MM & Fx & CAVE & MCI \\
Yu et al. 2025~\cite{yu2025} & 1 & H & N & MM & Fx & HMD & MCI \\
Mirmiran et al. 2026~\cite{mirmiran2026} & 2 & G & S & MM & M & Desktop & Dem \\
Goumopoulos et al. 2026~\cite{goumopoulos2026} & 2 & H & N & MM & A & Tablet & MCI \\
Stramba-Badiale et al. 2025~\cite{strambabadiale2025} & 1 & T & N & MM & A & Desktop & MCI \\
Park 2022~\cite{park2022b} & 1 & T & N & V & A & Desktop & MCI \\
Zygouris et al. 2022~\cite{zygouris2022} & 2 & G & N & V & Fx & Tablet & MCI/SCD \\
Li et al. 2024~\cite{li2024} & 1 & H & F & MM & M & HMD & MCI/Dem \\
Jha et al. 2020~\cite{jha2020} & 2 & G & F & MM & A & HMD & SCD \\
Rings et al. 2020~\cite{rings2020b} & 2 & H & F & MM & M & HMD & MCI/Dem \\
Burdea et al. 2015~\cite{burdea2015} & 2 & G & N & MM & A & Desktop & Dem \\
Chen et al. 2023~\cite{chen2023} & 2 & G & F & V & M & HMD & MCI/Dem \\
\bottomrule
\end{tabular}
\par\smallskip
{\footnotesize\raggedright
\textbf{T}: tier (1\,=\,controlled efficacy, 2\,=\,design/feasibility).\quad
\textbf{Par}: G\,=\,game-based, T\,=\,training-based, H\,=\,hybrid.\quad
\textbf{Imm}: F\,=\,fully immersive, S\,=\,semi-immersive, N\,=\,non-immersive.\quad
\textbf{FB}: MM\,=\,multimodal, V\,=\,visual only.\quad
\textbf{Adj}: A\,=\,automatic/performance-adaptive, M\,=\,manual/therapist-set/graded, Fx\,=\,fixed/none (\S\ref{sec:tax-adaptation}).\quad
\textbf{HW}: HMD, Desktop, Tablet, CAVE, Proj\,=\,projection, Kinect.\quad
\textbf{Pop}: MCI\,=\,mild cognitive impairment, Dem\,=\,dementia, AD\,=\,Alzheimer's disease, SCD\,=\,subjective cognitive decline (``/'' denotes a mixed sample).\quad
The two Park entries~\cite{park2022, park2022b} are separate RCTs of the same system.\par}
\end{table*}

\subsection{Design Paradigm}
\label{sec:tax-paradigm}
Out of 34 systems, only 20 were classified as a single paradigm. Of the 20 systems, 8 were identified as game-based systems, while 12 were training-based systems. The remaining 14 (41\%) consisted of both approaches, which makes the hybrid category the largest. The convergence came from both ways. There was an inclusion of different game mechanics such as scoring, levels and reward mechanics in training-based systems. Similarly, game-based systems also included structured and clinical tasks such as spatial navigation or memory training tasks. These findings indicate that paradigm labels reflect the primary design emphasis of a system rather than different intervention categories. Since nearly half of the systems do not fit the usual two categories, it is clear that this binary model does not match how the field actually looks.

\subsection{Immersion Level}
\label{sec:tax-immersion}
A fully immersive system was the most dominant one and used a head-mounted display in 19 studies out of 34. In contrast, only a smaller group was non-immersive (11) and used different technologies such as desktop, tablet and monitor. Additionally, 4 systems provided a semi-immersive experience through CAVE or projection technologies. Immersion level was not associated with intervention paradigm. Both game-based and training-based paradigms included all types of immersion. For example, among training-based systems, 6 were fully immersive, 4 were non-immersive and 2 were semi-immersive. These patterns were distributed similarly across game-based systems.

\subsection{Feedback Modality}
\label{sec:tax-feedback}
Most systems used multimodal feedback. In total, 29 out of 34 studies used visual and auditory channels. Some studies included touch, body-position and airflow cues. Only 5 studies were found to use visual feedback alone. Since a multimodal feedback system is used universally across both paradigms, this axis does not differentiate the paradigms.

\subsection{Difficulty Adaptation}
\label{sec:tax-adaptation}
There was a similarity in difficulty management across all studies. Out of 34 studies, 27 adjusted task difficulty, while 7 of them either kept it fixed or did not report it. Of the 27, 13 used automatic and performance-driven adaptation. This adaptation was done through machine learning and an LLM mechanism. The remaining 14 studies used manual adaptation. The rules here were mostly set by a therapist. Automatic adaptation was found in all paradigms such as game-based (4 of 8), training-based (4 of 12) and hybrid (5 of 14). This means the paradigms are not distinguishable by adaptivity. 

\subsection{Hardware}
\label{sec:tax-hardware}
The most common platform was head-mounted displays (HMDs) which was used in 21 studies out of 34. There were different devices used such as Oculus/Meta Quest, HTC Vive, Pico, Samsung Odyssey, Valve Index and Fove. The rest of 13 studies included non-HMD devices. There were 5 desktop computers, 4 CAVE or projection environments, 3 tablets and 1 Kinect-plus-screen setup. The paradigm difference was not marked by HMD usage since they were evenly distributed in game-based, training-based and hybrid systems. 

\subsection{Statistical Evaluation of Paradigm Differences}
\label{sec:tax-synthesis}
A cross-tabulation showed overlap between game-based, training-based and hybrid systems. To examine the overlap, we measured the association between interventions and each design feature using the Fisher-Freeman-Halton test. There were no significant connections found for immersion ($p=.96$), difficulty adaptation ($p=.40$), hardware ($p=.61$) or feedback ($p=.17$). However, there were tendencies in hybrid systems to utilize multimodal feedback more often, though it was not statistically significant. The analysis was further conducted using two pure paradigms, game-based and training-based systems only. No significant differences were found for immersion ($p=1.00$), adaptation ($p=.73$), feedback ($p=1.00$) and hardware ($p=.66$) either.

The equivalence between the two paradigms could not be demonstrated due to the small sample. However, the results indicate that there is no significant technological differentiation between them. The findings are also supported by the descriptive results. They showed considerable overlap across the design features and identified hybrid systems as the most common category (14 of 34 systems). Ultimately, the differentiation between the game-based and training-based systems arises from design emphasis not from any technological differences. This point is very important for interpreting past reviews (\S~\ref{sec:outcomes}). Comparing ``Game vs Training'' often includes similar technology. This finding is a potential reason behind the inconsistent reports in the literature.

Our five dimensions describe how a system is delivered rather than the experiential features. Those features include reward structure, narrative, challenge--skill balance and repetition. Two systems can look identical on our axes but they might differ a lot based on these features. Therefore, our results show only that the label does not predict technological setup. They do not show that the two paradigms are equivalent. The current literature does not report the experiential features in enough detail. That lack of detail is a part of the reporting gap discussed in \S~\ref{sec:vrres-results}.

This gap matters. Two tasks can make the same cognitive demand and still feel very different to the user. One may be engaging and the other may be dull. This difference affects how well people stick with the training. The included studies do not describe their tasks in enough detail for us to compare this. These limits restrict the taxonomy's usefulness. A future analysis that includes reward schedules, narrative features and challenge settings could reveal meaningful differences between the two paradigms. At present, the literature does not contain enough information to support such a comparison.

\section{Cognitive Effects of Controlled Interventions (RQ2)}
\label{sec:outcomes}
We combined the cognitive results from the 16 Tier-1 studies. Since there were variations in the outcome measures, comparison groups and reporting styles, we addressed them through a narrative approach. Effect sizes were mentioned only when they were reported in the original study. Table~\ref{tab:outcomes} shows each study's design and main cognitive result. Although the studies differed in their intervention designs, participant characteristics, and assessment methods, the analysis enabled the identification of broad patterns in reported cognitive outcomes.

\subsection{Overall Direction of Effect}
\label{sec:out-direction}
Most of the studies (13 out of 16) found that VR outperformed the comparator on at least one cognitive outcome. The studies~\cite{buele2024,mrakicsposta2018,strambabadiale2025} which showed exceptions include the two smallest in the corpus (N = 10 and 18). Yet these studies did not show any negative result. There was either an improvement within the VR group or a positive trend. This overall result is consistently positive. However, the size of the benefit was not even. This variation appears to track the comparator rather than the paradigm. Studies with no-treatment or usual-care comparators~\cite{kwan2024,hsieh2018,li2024,park2022,ip2025} reported larger advantages, whereas studies with active cognitive comparators~\cite{buele2024,park2022b,strambabadiale2025,kua2025} reported smaller gains. This pattern is only a descriptive observation based on a small and varied set of studies, and comparator type overlaps with sample size and outcome choice.

Intervention dose varied a lot across the Tier~1 studies. Session counts ranged from 8 to about 48. Session duration ran from 20 to 100 minutes. Total contact time ranged from about 4 hours~\cite{strambabadiale2025} to about 48 hours~\cite{hsieh2018}, with a median near 14 hours. Two studies~\cite{kua2025,li2024} did not report enough detail to work this out. Dose is not independent of the other study features. The three studies with null or marginal results~\cite{buele2024,mrakicsposta2018,strambabadiale2025} were positioned at the low end. In contrast, the two largest doses~\cite{hsieh2018,thapa2020} both used comparators with little cognitive content. Contact was often unmatched between arms as well. In Kua~\cite{kua2025} the VR group attended ten supervised weekly sessions while controls were reviewed once a month with homework tasks. Therefore, efficacy differences cannot be explained by design alone, because dose and comparator type vary together across the corpus.

Effect sizes were reported unevenly. Six reported Cohen's $d$, with values from about 0.3 to 1.9~\cite{hsieh2018,kua2025,yu2025,strambabadiale2025,liao2019,buele2024}. Five reported partial $\eta^2$ where values range from 0.09 to 0.78~\cite{park2022,park2022b,li2024,thapa2020,yang2022}. The two measures cannot be pooled or compared directly. Hence, we do not report a combined average. The remaining studies~\cite{kwan2024,liao2020,baldimtsi2023,ip2025,mrakicsposta2018} gave only $p$ values. Most effects are moderate to large. But the biggest appear in domain-specific tests. Global scales show smaller changes. As a result, effect size depends on which outcome was reported.

\subsection{Effects by Cognitive Domain}
\label{sec:out-domain}
The benefit is not identical across all areas. The most consistent improvement is found in executive function. There were gains on tests like TMT, Stroop or the Frontal Assessment Battery~\cite{thapa2020,liao2019,ip2025,yang2022,kua2025,hsieh2018}. The largest effects are in spatial and visuospatial memory, specifically in the systems that were built around navigation or construction tasks. Park's waitlist-controlled trial~\cite{park2022} showed a large effect ($\eta^2=.67$). The active-controlled trial~\cite{park2022b} also showed a large effect ($\eta^2=.50$). Kua's figure-copy gain~\cite{kua2025} was also large ($d=1.2$). Stramba-Badiale et al.~\cite{strambabadiale2025} found a Corsi-backward effect. In contrast, global composite scores (MMSE, MoCA) are often insensitive in measuring these effects. They were non-significant in Thapa, Kua and Stramba-Badiale~\cite{thapa2020,kua2025,strambabadiale2025}. However, they were reported as positive results in Li, Kwan and Baldimtsi~\cite{li2024,kwan2024,baldimtsi2023}. Whether an effect appears depends on whether domain-specific outcomes were used. 

These results can be read in terms of near and far transfer. The most consistent gains came from near transfer. These were tests that resembled the trained activity. Spatial navigation training produced gains on spatial cognition~\cite{park2022,park2022b}. Besides, visual building tasks produced gains on figure copy~\cite{kua2025}. Executive training produced gains on the Trail Making Test~\cite{thapa2020,liao2019,ip2025}. Far transfer to global cognition was less reliable. MMSE and MoCA showed no significant between-group difference in several studies~\cite{thapa2020,kua2025,strambabadiale2025}. However, others reported improvement on global scales~\cite{kwan2024,li2024,yang2022,baldimtsi2023}. Studies showing far transfer effects were mostly those that used no treatment or usual care comparators. They also used longer interventions. This means the reported far transfer may reflect weak comparison conditions as much as true generalization.

\begin{table*}[t]
\caption{Cognitive outcomes of the 16 Tier~1 controlled studies.}
\label{tab:outcomes}
\footnotesize
\setlength{\tabcolsep}{4pt}
\begin{tabular}{@{}llrcp{0.26\textwidth}p{0.30\textwidth}@{}}
\toprule
\textbf{Study} & \textbf{Par} & \textbf{N} & \textbf{Comp} &
\textbf{Comparator} & \textbf{Principal cognitive result} \\
\midrule
Kwan et al. 2024~\cite{kwan2024} & H & 293 & UC & Waitlist usual care & $\uparrow$ global cognition; $\downarrow$ frailty; high adherence \\
Liao et al. 2020~\cite{liao2020} & T & 34 & A-M & Combined phys.\,+\,cog.\ (non-VR) & $\uparrow$ global cognition, $\uparrow$ delayed verbal recall, $\uparrow$ IADL \\
Buele et al. 2024~\cite{buele2024} & T & 34 & A-C & Pencil-and-paper cog.\ training & NS between groups (both improved); VR higher completion \\
Baldimtsi et al. 2023~\cite{baldimtsi2023} & H & 122 & A-P\,/\,A-M\,/\,NT & Bike; phys.\ exercise; mixed phys.\,+\,cog.; non-contact & $\uparrow$ global cognition vs.\ bike; memory maintained vs.\ decline \\
Ip et al. 2025~\cite{ip2025} & H & 55 & UC & Ordinary centre-based social care & $\uparrow$ cognition ($p{=}.008$), $\uparrow$ EF (TMT-A/B) \\
Mrakic-Sposta et al. 2018~\cite{mrakicsposta2018} & H & 10 & NT & No-treatment & NS (underpowered); $\downarrow$ oxidative stress \\
Hsieh et al. 2018~\cite{hsieh2018} & H & 60 & NT & No treatment (usual daily activity) & $\uparrow$ abstract thinking/judgment ($d{=}0.5$--$1.0$) \\
Liao et al. 2019~\cite{liao2019} & H & 34 & A-M & Active (phys.\,+\,cog.) & $\uparrow$ TMT-B, $\uparrow$ cognitive dual-task gait (borderline) \\
Thapa et al. 2020~\cite{thapa2020} & G & 66 & A-E & Active (health education) & $\uparrow$ EF (TMT-B), $\uparrow$ SDST; MMSE NS \\
Kua et al. 2025~\cite{kua2025} & T & 22 & A-C & Active (TAU\,+\,cog.\ training) & $\uparrow$ figure copy ($d{=}1.2$), $\uparrow$ FAB ($d{=}1.0$); MoCA NS \\
Park 2022~\cite{park2022} & T & 56 & NT & Waitlist & $\uparrow$ spatial cognition ($\eta^2{=}.67$), $\uparrow$ recall ($\eta^2{=}.09$) \\
Yang et al. 2022~\cite{yang2022} & H & 99 & A-P\,/\,NT & Exercise; passive control & $\uparrow$ MMSE, $\uparrow$ TMT-A; cognition favoured VR \\
Yu et al. 2025~\cite{yu2025} & H & 92 & A-M\,/\,A-E & Cog.\,+\,Tai-chi; health ed.\ & Superior to both comparators (large effect sizes) \\
Stramba-Badiale et al. 2025~\cite{strambabadiale2025} & T & 18 & A-C & TAU\,+\,non-embodied VR & $\uparrow$ Corsi-backward (marginal, $p{=}.074$); else NS \\
Park 2022~\cite{park2022b} & T & 50 & A-C & 2D computerized spatial task & $\uparrow$ spatial cognition ($\eta^2{=}.50$), $\uparrow$ memory ($\eta^2{=}.11$) \\
Li et al. 2024~\cite{li2024} & H & 60 & UC & Usual care & $\uparrow$ global cognition (CASI $\eta^2{=}.78$, MMSE, MoCA) \\
\bottomrule
\end{tabular}
\par\smallskip
{\footnotesize\raggedright
\textbf{Par}: paradigm (see Table~\ref{tab:taxonomy}).\quad
\textbf{Comp}: comparator type, coded per control arm.
NT\,=\,no treatment or waitlist (no added contact or services);
UC\,=\,usual care / treatment-as-usual (ongoing services continue);
A-C\,=\,active cognitive (structured cognitive training);
A-P\,=\,active physical (exercise or motor training without a cognitive component);
A-E\,=\,attention or education control (contact-matched, low cognitive demand);
A-M\,=\,active mixed (combined physical and cognitive).
Studies with more than one control arm carry one code per arm, separated by ``/''.\quad
$\uparrow$\,=\,VR superior; NS\,=\,no significant between-group difference.\quad
Effect sizes ($d$, $\eta^2$) are reported only where the source study reported them.\quad
\par}
\end{table*}

\subsection{Efficacy across Different Paradigms}
\label{sec:out-paradigm}
The studies cannot support a clear comparison between game-based and training-based systems for two reasons. First, Tier-1 is not balanced. There is only 1 study which is purely game-based~\cite{thapa2020}. In contrast, there are six training-based studies and nine hybrid studies. Second, the paradigms are not technologically different. Therefore, the differences in efficacy could not be explained by the paradigm labels. The largest effects came from training-based systems with large spatial gains~\cite{park2022,park2022b} and from a hybrid system ($\eta^2=.78$)~\cite{li2024}. The game-based study also showed positive results. Effect size does not follow the paradigm. The factors which explain efficacy are methodological such as comparator type (passive vs active), outcome choice (domain-specific vs global) and sample size. Consequently, the inconsistencies across prior reviews can be better explained by design and measurement differences rather than the game vs training paradigm.

Passive comparators give a weaker test of efficacy. Consequently, we also looked at the ten studies with an active comparator on their own~\cite{liao2020,buele2024,baldimtsi2023,liao2019,thapa2020,kua2025,yang2022,yu2025,strambabadiale2025,park2022b}. As expected, the effects here were smaller. But the pattern across paradigms stayed similar. Positive results came from training-based~\cite{liao2020,kua2025,park2022b}, hybrid~\cite{baldimtsi2023,liao2019,yang2022,yu2025} and game-based~\cite{thapa2020} systems alike. Both null results~\cite{buele2024,strambabadiale2025} were training-based studies. They had the smallest doses and the closest-matched comparators. Although the stricter test reduces the effect size, it does not alter the conclusion. Outcomes do not cluster by paradigm.

\subsection{Understanding the disagreements in Previous Reviews}
\label{sec:out-Understand}
Our findings help explain the inconsistencies across the prior reviews discussed in \S~\ref{sec:related}. Yuan et al.~\cite{yuan2025review} and Moulaei et al.~\cite{moulaei2024review} said game-based VR yields better cognitive outcomes. But our analysis indicates this is not directly related to the paradigm. There are two reasons behind that. First, these paradigms are not technologically different (\S~\ref{sec:tax-synthesis}). They are both the same with respect to immersion, feedback, adaptation and hardware. So the advantage of one paradigm should not be connected with any design feature. Second, the size of the benefit depends more on the comparator than on the paradigm itself (\S~\ref{sec:out-direction}). We have found that passive and waitlist controls show the largest effects. \\
The meta-analyses showing an advantage for game-based systems do not contradict our findings. Their advantages might be connected with other factors such as testing the game-based system often against weak comparators or including strong design features like rich feedback or adaptive difficulty. In such cases, the advantages come from the design or comparator rather than the paradigm itself. Therefore, we do not suggest choosing a winning paradigm. Rather, focus should be given to the design features and trial quality directly. This leads to an evidence-quality check in the next section.

\section{VR-RES: An Evidence-Quality Framework (RQ3)}
\label{sec:vrres-results}
Existing evaluation frameworks such as the Cochrane risk-of-bias tool RoB~2~\cite{sterne2019rob2}, the Downs and Black checklist~\cite{downsblack1998}, the PEDro scale and the GRADE approach~\cite{guyatt2008grade} are designed to check bias and certainty in clinical trials in general. However, these tools are not sufficient for evaluating VR cognitive rehabilitation for three reasons. First, they give little attention to technological reproducibility. Consequently, a study can score well on RoB~2 despite not mentioning hardware, software version, interaction type, or difficulty adaptation logic. But these details are necessary to reproduce or compare the interventions. Second, follow-up assessment is very crucial for assessing the effectiveness of a VR cognitive rehabilitation process. Since the disease causes a gradual decline in cognitive abilities, it is important to determine whether the treatment effect lasts over time. Yet, a trial without a retention checkup can score well on the traditional framework. Third, they do not consider demographic moderator reporting during the assessment. This is very important since effectiveness can vary based on age, education, sex or baseline severity.

Therefore, we introduce VR-RES (VR Rehabilitation Evidence Score), an evaluation tool with seven dimensions. This judges each Tier-1 study as 0, 1, or 2 per dimension, where the maximum total score is 14. We are not considering it as a replacement for the existing tools. Rather, we believe it can be effective in complementing them. It includes the bias-related features that those tools evaluate such as design rigor, blinding, and sample size. In addition, it also adds the dimensions that they leave out, which include reproducibility, follow-up and demographic reporting. Therefore, it is a transparent and criterion-based rubric that shows the systematic gaps that the existing tools do not consider. The dimensions and their scoring criteria are listed in Table~\ref{tab:vrres}.

The 16 Tier-1 studies give a mean score of 8.1 out of 14, where their range is 6 to 10 after applying VR-RES. Table~\ref{tab:vrres-scores} shows the results for each study. Two things are noticeable from the table. First, all studies score between 6 and 10, which means that prior studies are mainly of medium quality rather than a mix of strong and weak work. The second is that the same dimensions are missing in almost every study, which prevents them from scoring highly. 

The total score describes the corpus. It is not a tool for ranking or weighting studies. We do not use it to drop studies or change how we read any single result. The seven dimensions assess different things. Hence, adding them with equal weights is just a way of summarizing. It does not mean one dimension can make up for another. The dimension scores are the useful part. They show which gaps keep recurring across the field. We also checked whether the total tracks outcomes. We found that it does not. Studies with a positive effect averaged 8.3 whereas null studies averaged 7.0. The difference is driven by the two null studies that scored zero on sample size.

The foundations of the field are solid. All studies used standardized cognitive instruments (outcome validity = 2.0 across the board). Here, 14 out of 16 studies went for a randomized controlled design (design rigor mean = 1.88). Sample sizes met our upper anchor in most of the studies (mean = 1.44; nine studies had more than 50 participants). Where the evidence base falls short, it does so collectively and predictably.

The single most striking gap is follow-up assessment. Only 1 out of 16 studies~\cite{ip2025} measured cognitive outcomes after the intervention ended. It included a follow-up assessment after three months. The remaining 15 studies reported post-test outcomes only. Therefore, whether the cognitive gains are sustained over time is an unresolved issue for a clinical population that is characterized by progressive decline.

Blinding also shows weak results. Half of the studies (8 out of 16) had no blinding at all. Only four used double or assessor blinding (mean = 0.75). This makes most results vulnerable to bias from expectations and judgments.

Reproducibility was poor as well. Not a single study reported both protocol and software data. Four studies shared nothing at all (mean = 0.75). Since this is a tech-based treatment, replication and comparison are very difficult without access to the system~\cite{anjum2025sok}. This directly aligns with our RQ1 finding: paradigms cannot be confirmed if we do not have any access to the systems.

\subsection{Demographic Factors}
\label{sec:demographics}
All 16 Tier-1 studies reported basic demographic moderators such as age, sex and education. However, only 2 studies~\cite{buele2024,yu2025} actually analyzed outcomes by demographic subgroup (mean = 1.12). Age, sex and education are very important factors in both technology and cognitive aging. Yet, prior studies only recorded them in their studies without considering them as effect modifiers. However, interaction tests need larger samples than main-effect tests. Most of these studies were too small to support them. Their absence limits what we can learn about differences in treatment response. Yet, it does not weaken the main efficacy findings. Therefore, the field still cannot identify which participants benefit most from VR cognitive rehabilitation. This also makes RQ2 harder to interpret because outcomes do not vary by design type alone. Participant characteristics may also explain some of the variation. But since so few studies examined demographic factors, this explanation cannot be evaluated.

These factors matter more for VR than for conventional training. Someone who is using a headset for the first time spends the early sessions learning the interface rather than doing the task. Hence, their scores partly reflect technology familiarity. Socioeconomic background influences baseline cognitive reserve.
It also affects the ability to keep practicing outside the sessions. None of the included studies reported prior technology experience. Only one~\cite{yu2025} reported socioeconomic status. However, it was recorded as a baseline characteristic rather than examined as a moderator.

\begin{table}[t]
\caption{VR-RES dimensions and scoring rubric (0/1/2; max 14).}
\label{tab:vrres}
\small
\begin{tabular}{@{}p{0.27\columnwidth}p{0.65\columnwidth}@{}}
\toprule
\textbf{Dimension} & \textbf{0 / 1 / 2 anchors} \\
\midrule
Design rigor & case/pilot / non-randomized controlled / RCT \\
Sample size & $N<20$ / $N$ 20--50 / $N>50$ \\
Blinding & none / single-blind / double or assessor-blind \\
Outcome validity & ad-hoc / mixed / validated standardized instruments \\
Follow-up & post-test only / short / $\ge$3-month retention \\
Reproducibility & nothing shared / protocol / protocol + software/data \\
Demographic reporting and generalizability & pooled only / reported by group with baseline balance or covariate adjustment / prespecified moderator analysis \\
\bottomrule
\end{tabular}
\par\smallskip
{\footnotesize\raggedright
Anchors follow common benchmarks rather than a formal derivation. At $\alpha=.05$, a two-group study with $N=20$ has about 40\% power for a large effect and $N=50$ about 79\%. This matches the effect sizes most of these studies report. Study design still affects power. A big study divided across many arms may not have enough participants, but a smaller study might. A follow-up gap under three months makes it hard to tell a lasting gain from a leftover practice effect. For blinding, participants always know they are wearing a headset. Hence, assessor blinding is as far as VR trials can go, which is why we score it at the top. Our blinding scores should therefore not be read against those from drug trials.\par}
\end{table}

\begin{table*}[t]
\caption{VR-RES scores for the 16 Tier~1 studies.}
\label{tab:vrres-scores}
\footnotesize
\setlength{\tabcolsep}{4pt}
\begin{tabular}{@{}lcccccccc@{}}
\toprule
\textbf{Study} & \textbf{Des} & \textbf{Smp} & \textbf{Bln} & \textbf{OV} & \textbf{FU} & \textbf{Rep} & \textbf{Dem} & \textbf{Tot} \\
\midrule
Kwan et al. 2024~\cite{kwan2024} & 2 & 2 & 1 & 2 & 0 & 1 & 1 & 9 \\
Liao et al. 2020~\cite{liao2020} & 2 & 1 & 1 & 2 & 0 & 1 & 1 & 8 \\
Buele et al. 2024~\cite{buele2024} & 2 & 1 & 1 & 2 & 0 & 1 & 2 & 9 \\
Baldimtsi et al. 2023~\cite{baldimtsi2023} & 2 & 2 & 0 & 2 & 0 & 1 & 1 & 8 \\
Ip et al. 2025~\cite{ip2025} & 1 & 2 & 0 & 2 & 2 & 1 & 1 & 9 \\
Mrakic-Sposta et al. 2018~\cite{mrakicsposta2018} & 2 & 0 & 0 & 2 & 0 & 1 & 1 & 6 \\
Hsieh et al. 2018~\cite{hsieh2018} & 1 & 2 & 0 & 2 & 0 & 1 & 1 & 7 \\
Liao et al. 2019~\cite{liao2019} & 2 & 1 & 2 & 2 & 0 & 1 & 1 & 9 \\
Thapa et al. 2020~\cite{thapa2020} & 2 & 2 & 0 & 2 & 0 & 1 & 1 & 8 \\
Kua et al. 2025~\cite{kua2025} & 2 & 1 & 2 & 2 & 0 & 0 & 1 & 8 \\
Park 2022~\cite{park2022} & 2 & 2 & 2 & 2 & 0 & 1 & 1 & 10 \\
Yang et al. 2022~\cite{yang2022} & 2 & 2 & 0 & 2 & 0 & 0 & 1 & 7 \\
Yu et al. 2025~\cite{yu2025} & 2 & 2 & 1 & 2 & 0 & 1 & 2 & 10 \\
Stramba-Badiale et al. 2025~\cite{strambabadiale2025} & 2 & 0 & 0 & 2 & 0 & 1 & 1 & 6 \\
Park 2022~\cite{park2022b} & 2 & 1 & 2 & 2 & 0 & 0 & 1 & 8 \\
Li et al. 2024~\cite{li2024} & 2 & 2 & 0 & 2 & 0 & 0 & 1 & 7 \\
\midrule
\textit{Mean} & \textit{1.88} & \textit{1.44} & \textit{0.75} & \textit{2.00} & \textit{0.12} & \textit{0.75} & \textit{1.12} & \textit{8.06} \\
\bottomrule
\end{tabular}
\par\smallskip
{\footnotesize\raggedright \textbf{Des}\,=\,design rigor, \textbf{Smp}\,=\,sample
size, \textbf{Bln}\,=\,blinding, \textbf{OV}\,=\,outcome validity,
\textbf{FU}\,=\,follow-up, \textbf{Rep}\,=\,reproducibility,
\textbf{Dem}\,=\,demographic reporting, \textbf{Tot}\,=\,total (max 14); scoring
anchors in Table~\ref{tab:vrres}.\par}
\end{table*}

\section{Discussion}
\label{sec:gaps}
\subsection{Principal Findings}
In this study, we analyzed a key assumption: whether game-based and training-based VR systems differ technologically in cognitive rehabilitation. Through our systematization of 34 recent works, we did not find much to support this assumption. There are no significant differences in immersion, feedback, difficulty adaptation, or hardware (\S\ref{sec:tax-synthesis}) between the paradigms. These findings suggest that game-based and training-based differentiation arises from design emphasis rather than the technological features we coded. The interpretation is also supported by the analysis of cognitive outcomes (\S\ref{sec:outcomes}) of recent studies. Here, the positive cognitive effects of the studies are influenced more by factors such as comparator type, outcome measures and sample size rather than the paradigms. The VR-RES assessment (\S\ref{sec:vrres-results}) also showed that the overall quality of evidence was moderate (mean score = 8.1/14). There are some common limitations reported in prior studies that include a lack of long-term follow-up, limited system reproducibility and rare analysis of demographic factors.

\subsection{Relation to Prior Reviews and Their Limitations}
These findings clarify the inconsistencies found in the prior studies. Yuan et al.~\cite{yuan2025review} and Moulaei et al.~\cite{moulaei2024review} found a cognitive advantage of game-based VR. In contrast, Yang et al.~\cite{yang2025jmirreview}, Santos et al.~\cite{santos2025review} and Gao et al.~\cite{gao2025review} found no consistent paradigm effect. There is a limitation found in all the prior studies. They did not evaluate whether the systems are genuinely different before comparing the game-based and training-based systems. Because the paradigms are not different (\S\ref{sec:tax-synthesis}), any paradigm-level advantage is associated with the design features and the trial conditions, as we discussed in \S\ref{sec:out-Understand}.

Two additional limitations are also found in the previous studies. First, these studies often evaluate the clinical effectiveness of these interventions, while the technological design of VR systems is rarely analyzed. Different kinds of important features such as immersion, feedback, adaptive difficulty and hardware, are often not analyzed systematically. As a result, it becomes difficult to understand the linkage of outcomes with particular design characteristics. Second, study quality is usually assessed by tools developed for clinical trials such as RoB 2~\cite{sterne2019rob2}, the Downs and Black checklist~\cite{downsblack1998} and GRADE~\cite{guyatt2008grade}. These frameworks do not cover important issues for software-based interventions such as system reproducibility, long-term effectiveness and performance across different populations.

The primary studies included in our study also reflect the same limitations. Only one controlled trial reported follow-up measurements~\cite{ip2025} after the interventions, and only two examined outcomes across demographic subgroups~\cite{buele2024,yu2025}. None of the systems shared a complete protocol and runnable system. Since these diseases are characterized by a gradual decline of cognitive abilities~\cite{petersen2004,livingston2020}, follow-up assessment is very crucial to assess whether the benefits are sustained over time. Similarly, the lack of reproducible specifications makes it difficult to compare with alternative VR approaches and to determine which one contributes to successful outcomes. 
\subsection{Applications for XR Research and Practice}
The main application for the VR community is that future research should focus on specific design features rather than broad paradigm labels. Different features like immersion, feedback and adaptive difficulty vary independently, having no prior linkage with game-based and training-based systems. Researchers should evaluate cognitive outcomes based on these features. Comparing broad paradigms may hide the influence of individual design choices. Feature-level comparisons are better at identifying what drives effectiveness. Another significant challenge is the lack of reproducibility. Without full configuration and an adaptive mechanism, a VR rehabilitation system cannot be compared or replicated. Hence, the ability to identify which design features are responsible for observed benefits becomes limited. Therefore, the recommendations in the next section should be read not only as guidance for future studies but also as reporting and development standards for VR rehabilitation systems.
\section{Recommendations}
\label{sec:recommendations}
\subsection{Study Design}

\subsubsection{Focus on features, not paradigms:} Game-based and training-based systems are not truly different in terms of technology (\S\ref{sec:taxonomy}). Consequently, more focus should be given to the specific design features like immersion, feedback modality, difficulty adaptation, and interaction type rather than comparing the two broad categories. The most important information can be retrieved through varying one design factor within a single system (e.g.\ adaptive vs.\ fixed difficulty) without comparing the entire paradigms against one another.
\subsubsection{Use active comparators:} The size of the effect is more connected with the comparator than the intervention itself (\S\ref{sec:out-direction}). Passive controls tend to provide exaggerated benefits. Therefore, researchers should consider a comparator that is equivalent in training dose and contact time to have a fair comparison.
\subsubsection{Plan sample size for real effects:} The two smallest studies (10 and 18 participants) failed to show clear results because they were too small. Hence, proper sample sizes should be considered to expect strong and reliable findings.

\subsection{Evaluation and Testing}

\subsubsection{Add follow-up testing:} This is the biggest gap found in this field. Of the 16 controlled studies, 15 did not include any follow-up assessment. Since the disease is characterized by progressive decline, it is important to reassess participants after 3--6 months to determine whether the benefits persist over time.
\subsubsection{Include relevant tests:} Use tests that closely align with the targeted skill. Sometimes general tests (MMSE or MoCA) miss small but important improvements (\S\ref{sec:out-domain}). Therefore, it is more reliable to define and preregister the primary outcome before starting the study.
\subsubsection{Analyze demographic moderators:} Age, sex, and education are usually recorded in studies, but they are almost never analyzed (\S\ref{sec:demographics}). To find out the most beneficial group, subgroup or moderator analyses should be planned in advance.
\subsubsection{Report tolerability and engagement:} Focus should also be given to determine whether people can tolerate VR, whether they stay engaged and whether there are problems like cybersickness, mental load, fatigue or loss of balance~\cite{mahmud2023feedbackmodalities, mahmud2026multimodalwalking, pavel2025vrfallnet}.

\subsection{Reporting and Reproducibility}

\subsubsection{Make the system reproducible:} No study shared both its protocol and software (\S\ref{sec:vrres-results}). Since the treatment is solely dependent on the system, there should be a clear mention of hardware, engine and version, interaction logic, difficulty adaptation logic and session dose. Researchers should also share the parameters or task definition to make the reproducibility possible. 
\subsubsection{Use a common framework for reporting:} The seven VR-RES features can be more than just an evaluation tool. They can also be considered as checklists for reporting a study. If future studies describe the systems and outcomes following this tool, it will be easier to compare different studies. In addition, this tool eliminates the risk of missing information.

\section{Limitations}
\label{sec:limitations}
There are some limitations in our study. Screening and data extraction were completed by a single reviewer, which may lead to the risk of selection and extraction errors. We applied predefined eligibility criteria and a structured extraction form in order to reduce the risk. The outcome measures, comparator groups, and reporting formats differ significantly in our included studies. To address these, we report the effects and effect sizes according to the original studies instead of meta-analysis. Furthermore, during the screening stage, we narrowed our scope to cognitive rehabilitation from a broad focus on motor and cognitive applications. This change was clearly documented in our study. The VR-RES framework was specifically developed for this review and has not gone through any external validation. Finally, the Tier-1 corpus is very small ($n=16$) and includes only one purely game-based study. Therefore, our findings do not support a clear relationship between the efficacy and intervention paradigm. At the same time, they should not be interpreted as proof that the two approaches produce identical outcomes.

\section{Conclusion}
\label{sec:conclusion}
We systematized 34 studies that investigated virtual reality for cognitive rehabilitation in mild cognitive impairment, dementia and Alzheimer's disease. Of these, 16 were controlled efficacy trials and 18 were design and feasibility studies. Here, we challenge the assumption that game-based and training-based systems use different technological approaches. Different features such as immersion, feedback, difficulty adaptation and hardware distribution were similar across both categories while almost half of the studies went for a hybrid approach. The controlled studies indicate a consistent cognitive improvement. However, these benefits are not related to the design paradigm. Instead, different methodological choices such as comparator selection, outcome measures and sample size play a crucial role in those observed effects. The VR-RES appraisal provides a clear explanation for the inconsistent findings in the current literature. Methodological quality in earlier studies was moderate, where longitudinal follow-up was nearly absent.
Limited attention was given to blinding and reproducibility. Although demographic moderators are often reported, they are rarely analyzed. These findings suggest that inconsistencies found in the earlier studies arise from variation in study design and outcome measurement. The findings show little relation to differences between game-based and training-based interventions. Therefore, in the future, more emphasis should be given to the comparison of design features, longer follow-up periods and standardized outcome measures. In addition, identifying the populations that got the highest benefit and reporting sufficient details are also essential to facilitate reproducibility. Therefore, VR-RES is introduced as a framework to guide a more systematic approach in this field.

\bibliographystyle{ACM-Reference-Format}
\bibliography{references}
\end{document}